\documentclass[11pt]{amsart}

\pdfoutput=1

\usepackage{amsmath,amssymb,amsfonts,amsthm}
\usepackage[margin=2cm]{geometry}
\usepackage{caption,subcaption}
\usepackage[colorlinks]{hyperref}
\usepackage{tikz}

\definecolor{coGB}{HTML}{1F77B4}
\definecolor{coWG}{HTML}{2CA02C}
\definecolor{coAR}{HTML}{D62728}

\hypersetup{
    colorlinks = True,
    linkcolor  = coAR,
    citecolor  = coWG,
    urlcolor   = coGB,
}

\newcommand{\xpos}{\boldsymbol{x}}
\newcommand{\ypos}{\boldsymbol{y}}

\begin{document}


\title{Synthetic aperture imaging of dispersive targets}


\author{Arnold D.~Kim and Chrysoula Tsogka}


\address{Department of Applied Mathematics, University of California,
  Merced, 5200 North Lake Road, Merced, CA 95343, USA}


\email{adkim@ucmerced.edu, ctsogka@ucmerced.edu}

\maketitle


\begin{abstract}
  We introduce a dispersive point target model based on scattering by
  a particle in the far-field.
  The synthetic aperture imaging problem is then expanded to identify
  these targets and recover their locations and frequency dependent
  reflectivities. We show that Kirchhoff migration (KM) is able to
  identify dispersive point targets in an imaging region. However, KM
  predicts target locations that are shifted in range from their true
  locations. We derive an estimate for this range shift for a single
  target. We also show that because of this range shift we cannot
  recover the complex-valued frequency dependent reflectivity, but we
  can recover its absolute value and hence the radar cross-section
  (RCS) of the target. Simulation results show that we can detect,
  recover the approximate location, and recover the RCS for dispersive
  point targets thereby opening opportunities to classifying important
  differences between multiple targets such as their sizes or material
  compositions.

  \bigskip

\end{abstract}

\section{Introduction}

Synthetic aperture radar is an imaging modality in which an airborne
antenna is used to collect the reflected signal from a region of
interest on the ground. High resolution images are reconstructed by
coherently processing the signals along the known flight trajectory
\cite{cheney2001mathematical,cummingwong2005, cheney2009fundamentals,
  moreira2013tutorial}.  These images provide an estimate of the
spatial dependence of the reflectivity often ignoring its frequency
content. However, the dispersive nature of the reflectivity of targets
is of great interest as it can be helpful for material identification,
for example.

A natural approach that has been proposed to that effect is based on
dividing the frequency band into sub-bands and then creating an image
for each sub-band \cite{Albaneseproc2012,
  albanesemedina2013}. Although the individual images have lower
resolution, they can be successfully used to provide information about
the frequency dependence of the reflectivity. In the same spirit,
frequency and direction dependent reflectivity has been successfully
reconstructed in \cite{borcea2016synthetic} using sparsity constraint
optimization approaches while dividing the bandwidth and the array
aperture in sub-bands and sub-apertures respectively. The reflectivity
in this case has a four dimensional parametrization, {\it i.e.,} space,
frequency and direction. Computational complexity limits the
applicability of this method for on-the-fly scenarios.
      
Exploiting Doppler shift in the SAR ambiguity function
\cite{cheney2001mathematical} has been extended to frequency dependent
reflectivities and an expression for the SAR point spread function in
space and frequency domain has been derived
\cite{cheneydoppler2013}. This point spread function allows for
reconstructing an image in which each pixel provides frequency
dependent information about the reflectivity
\cite{sotirelisproc2013}. The approach gives promising results but
achieving high range resolution remains a challenge.

Another way to account for dispersive targets is to consider the
signal in the time domain in which case the scattering delay induced
by the target needs to be separated from the propagation delay. This
is a challenging problem that has been addressed in
\cite{gilmantsynkov2019} provided the synthetic aperture is wide
enough. The important question of detectability of this scattering
delay in the presence of speckle has been evaluated using statistical
divergence measures in \cite{gilmantsynkov2021}.

In this paper we consider a realistic model for a frequency dependent
reflectivity and propose an imaging method based on coherent
back-projection. This method allows us to first image the spatial
location of the targets and then determine their frequency dependent
reflectivities. High resolution imaging of the target location is
obtained using the the tunable synthetic aperture radar imaging
approach of \cite{kim2022tunable}. This method relies on a simple
mathematical transformation of the classical SAR image depending on a
user-defined parameter, $\epsilon$. The resulting image scales the
traditional SAR image resolution by $\sqrt{\epsilon}$ thus achieving
sub-wavelength target localization.

Due to the frequency dependence of the reflectivity, the target's
location is reconstructed up to a shift in range. Our theoretical
analysis provides an estimate of this shift and shows that it is
inherently connected to fundamental scattering properties of the
target, namely the reflectivity. Once the target location is
estimated, the radar cross-section (RCS) of the target can be
recovered. Promising results are obtained for single and multiple
targets scenarios. Gaining access to RCS information is very important
for some remote sensing applications as it provides target
classification information in addition to detection and spatial
localization.

%
%
%

The remainder of this paper is as follows. In Section
\ref{sec:particle} we review the elementary theory of scattering by a
particle and use that to introduce our dispersive point target
model. In Section \ref{sec:sar} we describe the SAR imaging problem
for dispersive point targets. We apply Kirchhoff migration (KM) to
identify the location which, in turn, enables the recovery of the
radar cross-section for a single dispersive point target in Section
\ref{sec:single-target}. There we show that KM accurately identifies
the target location in cross-range, but may produce a shift in
range. We derive an estimate for this range shift that identifies the
key mechanism producing this shift. We extend these results for
multiple targets in Section \ref{sec:multiple-targets}. For that case,
we introduce an elementary linear regression problem to obtain the
radar cross-sections for each of the targets. In Section
\ref{sec:conclusions} we give our conclusions. Appendix
\ref{sec:appendix} gives a description of scalar wave scattering by a
sphere which we use to generate frequency dependent reflectivities
used in the simulations results shown here.

\section{Scattering by a particle}
\label{sec:particle}

We briefly review elementary aspects of scattering by a particle and
use that to introduce our dispersive point target model.  Consider the
observation of the scattered field $U_{s}$ at distance $R$ away from a
particle with $R > d^{2}/\lambda$ where $d$ is the particle diameter
and $\lambda$ the wavelength of the incident light. For this case, the
leading behavior of the scattered field is~\cite{ishimaru}
\begin{equation}
  U_{s} \sim f(\hat{\boldsymbol{o}},\hat{\boldsymbol{\imath}};\omega)
  \frac{e^{\mathrm{i} \omega R/c}}{R}
  \label{eq:far-field-scattering}
\end{equation}
where $\hat{\boldsymbol{o}}$ is the direction of observation,
$\hat{\boldsymbol{\imath}}$ is the propagation direction of the
incident plane wave, $\omega$ is the frequency and $c$ is the wave
speed. The leading behavior given by \eqref{eq:far-field-scattering}
is a spherical wave modified by $f$, the scattering amplitude. The
scattering amplitude contains the amplitude and phase of the scattered
field in the far-field at frequency $\omega$.

In synthetic aperture imaging, we measure only the backscattered field
corresponding to $\hat{\boldsymbol{o}} =
-\hat{\boldsymbol{\imath}}$. The radar cross-section (RCS),
\begin{equation}
  \sigma_{\text{RCS}}(\omega) = 4 \pi | f(-\hat{\boldsymbol{\imath}},
  \hat{\boldsymbol{\imath}}; \omega) |^{2}
\end{equation}
gives a measure of the power backscattered by the particle. The RCS as
a function of $\omega$ depends on the size, shape, and material
properties of the particle.

SAR imaging methods such as KM tend to produce images of general
objects that exhibit peaks at the most singular portions of those
objects, {\it e.g.}~closest boundaries, corners,
etc~\cite{cheney2009fundamentals}. For this reason, point target
models are commonly used for those imaging problems. The point target
model that is typically used for SAR imaging problems assumes that the
scattered field measured at a point $\xpos$ is given by
\begin{equation}
  U^{s}(\xpos) = \rho \frac{e^{\mathrm{i} \omega | \xpos - \ypos | /
      c}}{4 \pi | \xpos - \ypos |} U^{\text{inc}}(\ypos).
  \label{eq:point-target}
\end{equation}
Here, $U^{\text{inc}}$ is the incident field, $\ypos$ is the location
of the point target and $\rho$ is a complex scalar called the
reflectivity. Comparing \eqref{eq:point-target} with
\eqref{eq:far-field-scattering}, we see that the reflectivity is the
scattering amplitude when $f$ is assumed to be independent of
direction and frequency.

We introduce an extension to \eqref{eq:point-target} through inclusion
of a frequency dependent reflectivity $\varrho(\omega)$ according to
\begin{equation}
  U^{s}(\xpos) = \varrho(\omega) \frac{e^{\mathrm{i} \omega | \xpos -
      \ypos | / c}}{4 \pi | \xpos - \ypos |} U^{\text{inc}}(\ypos).
  \label{eq:dispersive-point-target}
\end{equation}
We call \eqref{eq:dispersive-point-target} the dispersive point target
model.  This model is characterized by the position $\ypos$ and the
frequency dependent reflectivity $\varrho(\omega)$.

In the numerical simulations that follow, we determine
$\varrho(\omega)$ from the scattering amplitude for a dielectric
sphere with radius $\alpha$ and relative refractive index $n_{rel}$
(see Appendix \ref{sec:appendix}). Using that reflectivity, we compute the RCS
through evaluation of
$\sigma_{\text{RCS}}(\omega) = 4 \pi | \varrho(\omega) |^{2}$.

\section{Synthetic aperture imaging}
\label{sec:sar}

In synthetic aperture radar (SAR) imaging, a single
transmitter/receiver is used to collect the scattered electromagnetic
field over a synthetic aperture that is created by a moving platform
\cite{cheney2001mathematical, cheney2009fundamentals,
  moreira2013tutorial}. The moving platform is used to create a suite
of experiments in which pulses are emitted and resulting echoes are
recorded by the transmitter/receiver at several locations along the
flight path. Let $f(t)$ denote the broadband pulse emitted and let
$d(s,t)$ denote the data recorded. Here, the measurements depend on
the slow time $s$ that parameterizes the flight path of the platform,
$\boldsymbol{r}(s)$, and the fast time $t$ in which the round-trip
travel time between the platform and the imaging scene on the ground
is measured.

\begin{figure}[htb]
  \centering
  \includegraphics[width=0.6\linewidth]{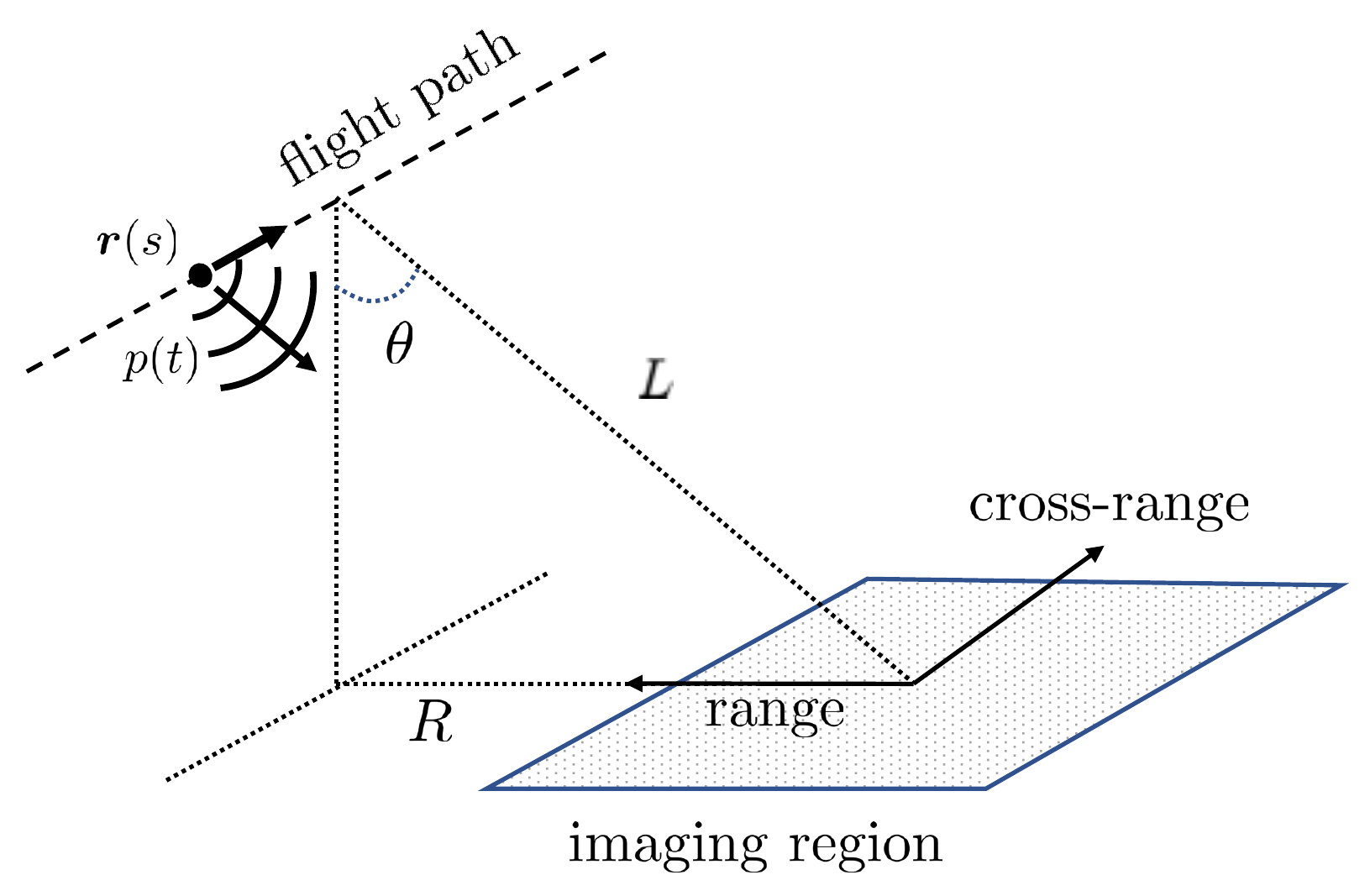}
  \caption{Synthetic aperture radar imaging schematic.}
  \label{fig:SAR-setup}
\end{figure}

High-resolution images of the probed scene can be obtained because the
data are coherently processed over a large synthetic aperture created
by the moving platform.  As illustrated in Fig.~\ref{fig:SAR-setup},
the platform is moving along a trajectory probing the imaging scene by
sending a pulse $p(t)$ and collecting the corresponding echoes. We
call range the direction that is obtained by projecting on the imaging
plane the vector that connects the center of the imaging region to the
central platform location. Cross-range is the direction that is
orthogonal to the range. Denoting the size of the synthetic aperture
by $a$ and the available bandwidth by $B$, the typical resolution of
the imaging system is $O((c/B) (L/R))$ in range and
$O(\lambda_{0} L/a)$ in cross-range. Here $c$ is the speed of light
and $\lambda_{0}$ the wavelength corresponding to the central
frequency while $L$ denotes the distance between the platform and the
imaging region and $R$ the offset in range.

In what follows, we use the start-stop approximation which neglects
displacements of the platform and targets in comparison with the
propagation of signals emitted and received on the platform. Let
$\xpos_{n}$ for $n = 1, \dots, N$ denote the positions of the
emitter/receiver along the flight path making up a synthetic
aperture. The imaging system operates with frequencies $\omega_{m}$
for $m = 1, \dots, M$ sampling the system bandwidth, $2 \pi B$.

Suppose there is dispersive point target located at $\ypos_{0}$, a
point in the imaging region, with frequency dependent reflectivity,
$\varrho_{0}(\omega)$. When the signal is emitted from the
emitter/receiver, it propagates into the medium, is incident on the
dispersive point target and scatters. The field scattered by the
dispersive point target is then measured on the emitter/receiver.  The
resulting measurement of the scattered field by the emitter/receiver
due to an isotropic and flat frequency point source at $\xpos_{n}$ is
\begin{equation}
  d_{mn} = \varrho_{0}(\omega_{m})
  \frac{e^{\mathrm{i} 2 \omega_{m} | \xpos_{n} - \ypos_{0}  | / c}}{(
    4\pi | \xpos_{n} - \ypos_{0} |)^{2}}, \quad
  m = 1, \dots, M, \quad
  n = 1, \dots, N.
  \label{eq:dmn}
\end{equation}
The matrix $D \in \mathbb{C}^{M \times N}$ whose entries are given by
\eqref{eq:dmn} contains the measurements.

The imaging problem is to determine the locations of targets and the
frequency dependent reflectivity in some specified imaging region. We
show below that we cannot recover the frequency dependent
reflectivity, in general. Instead we seek to recover the RCS for each
of the targets.

In the simulations results that follow, we use system parameters based
on the GOTCHA data set~\cite{casteel2007challenge}. In particular, we
have set $R = 3.55\, \text{km}$ and $H = 7.30\, \text{km}$, so that
$L = \sqrt{H^{2} + R^{2}} = 8.12\, \text{km}$.  The synthetic aperture
created by the linear flight path is $a = 0.13\, \text{km}$. The
central frequency is $\omega_0/(2\pi)=9.6\, \text{GHz}$ and the bandwidth is
$B = 622\, \text{MHz}$. Using $c = 3 \times 10^{8}\, \text{m/s}$, we
find that the central wavelength is $\lambda_{0} = 3.12\,
\text{cm}$. The imaging region is at the ground level $z = 0$. We use
$M = 25$ equi-spaced frequencies sampling the bandwidth, and $N = 32$
equi-spaced spatial measurements sampling the synthetic aperture.

\section{KM for a single dispersive point target}
\label{sec:single-target}

Let $\ypos$ denote a point in the imaging region.  We consider the
image formed through evaluation of the KM imaging function,
\begin{equation}
  I^{\text{KM}}(\ypos) = \sum_{m = 1}^{M} \sum_{n = 1}^{N} d_{mn}
  e^{-\mathrm{i} 2 \omega_{m} | \xpos_{n} - \ypos |/c}.
  \label{eq:IKM}
\end{equation}
Note that in \eqref{eq:IKM}, the entries of the data matrix are
back-propagated to $\ypos$ through multiplication by
$e^{-\mathrm{i} 2 \omega_{m} | \xpos_{n} - \ypos |/c}$.  Those results
are summed over spatial locations (sum in $n$) and frequencies (sum in
$m$). Through evaluation of this KM imaging function over a set of
points and plotting those results, we produce an image of targets in
the imaging region.

\begin{figure}[htb]
  \centering
  \includegraphics[width=0.48\linewidth]{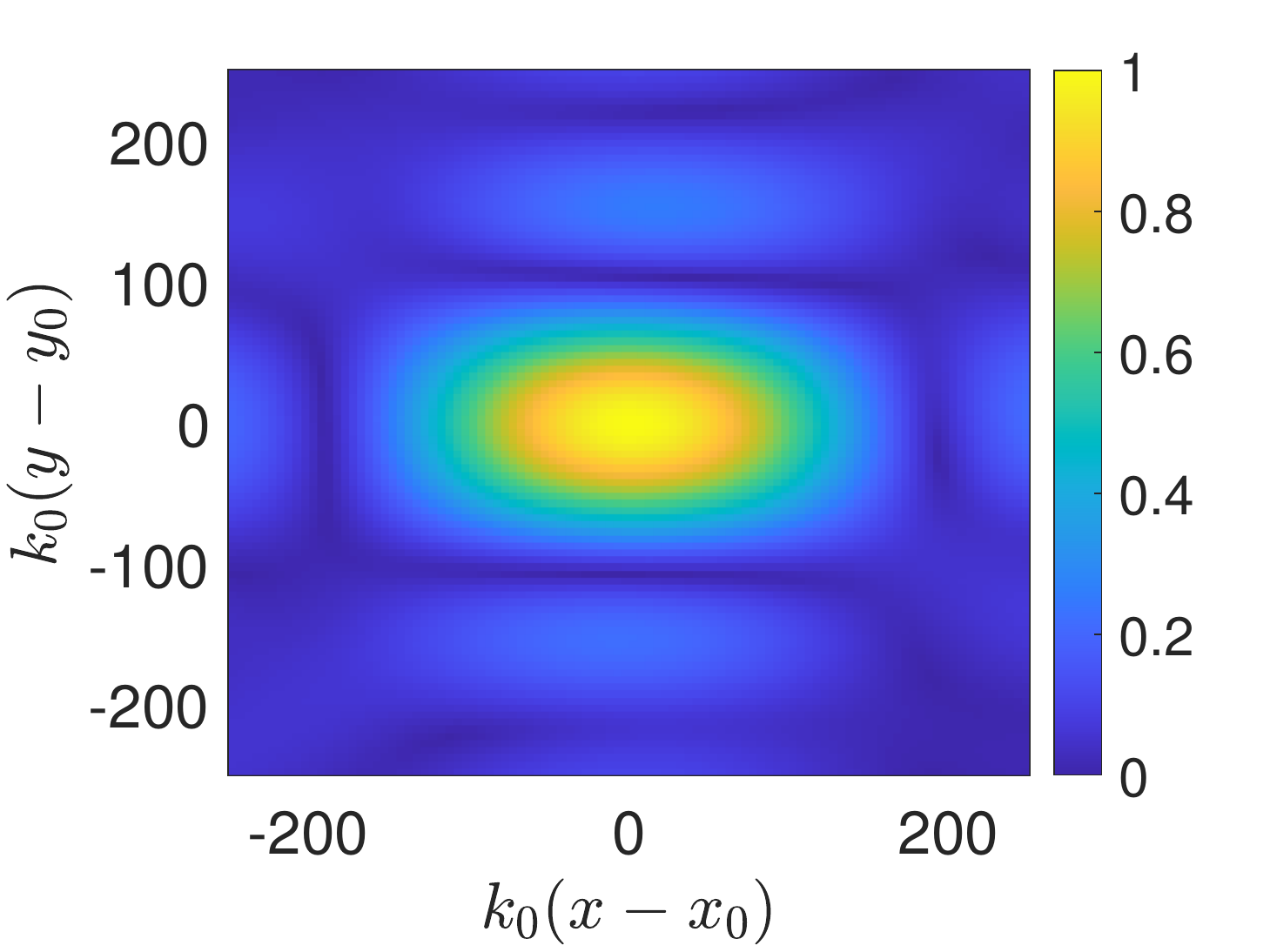}
  \caption{Evaluation of \eqref{eq:IKM} on an imaging region that is
    $500/k_{0} \times 500/k_{0}$ centered about the target location
    $(x_{0}, y_{0})$ with $k_{0}$ denoting the central wavenumber. The
    reflectivity was computed for a sphere with $k_{0} \alpha = 1.4$
    and $n_{rel} = 1.4$.  Measurement noise was added so that
    $\text{SNR} = 3.73$ dB.}
  \label{fig:2}
\end{figure}

In Fig.~\ref{fig:2} we show the result of evaluating \eqref{eq:IKM}
for a dispersive point target located at
$(k_{0} x_{0}, k_{0} y_{0}) = ( 273.713, -346.167 )$ over a
$500/k_{0} \times 500/k_{0}$ imaging region. The reflectivity was
computed for a sphere with $k_{0} \alpha = 1.4$ and $n_{rel} = 1.4$.
Measurement noise was added so that $\text{SNR} = 3.73$ dB. The image
shown in Fig.~\ref{fig:2} shows $I^{\text{KM}}$ normalized by its
maximum value. This image indicates the presence of a target through
its peak. The location of the peak predicts the location of the
target.  Away from the peak, we observe imaging artifacts as sidelobes
to the peak.

We have recently introduced a modification to KM that produces tunably
high-resolution images~\cite{kim2022tunable}. Let
$\tilde{I}^{\text{KM}}$ denote \eqref{eq:IKM} normalized by its
maximum as shown in Fig.~\ref{fig:2}. The modification to KM simply
requires evaluation of
\begin{equation}
  I_{\epsilon}^{\text{KM}} = \frac{\epsilon}{1 - (1 - \epsilon)
    \tilde{I}^{\text{KM}}},
  \label{eq:rKM}
\end{equation}
with $\epsilon$ denoting a user-defined parameter. The resolution of
the resulting image produced using \eqref{eq:rKM} scales with
$\sqrt{\epsilon}$.

\begin{figure}[htb]
  \centering
  \includegraphics[width=0.48\linewidth]{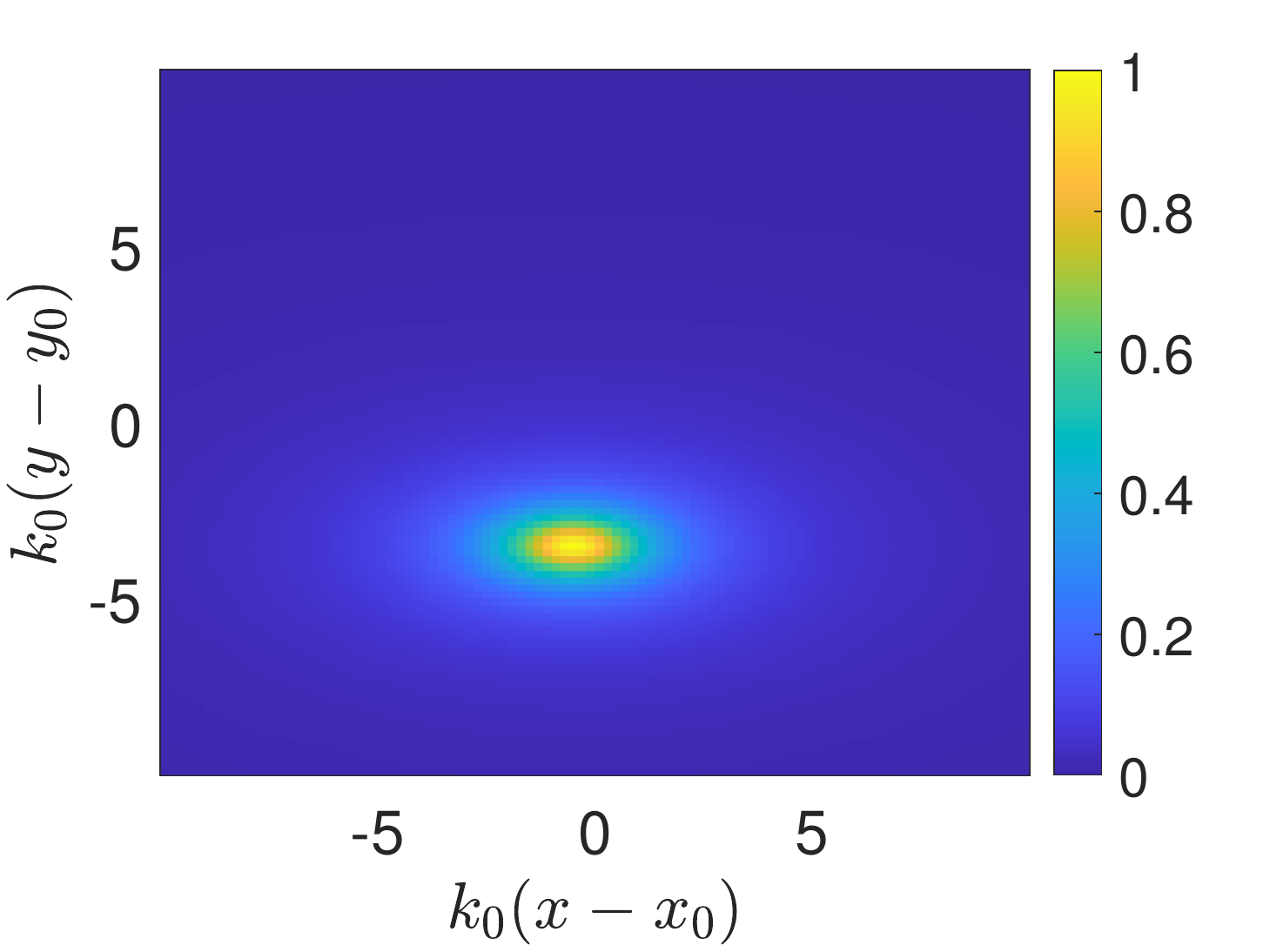}
  \caption{Evaluation of the modified KM given in \eqref{eq:rKM} with
    $\epsilon = 10^{-4}$ applied to the image shown in
    Fig.~\ref{fig:2} on an imaging region that is
    $20/k_{0} \times 20/k_{0}$ centered about the target location
    $(x_{0}, y_{0})$.}
  \label{fig:3}
\end{figure}

When we plot $I_{\epsilon}^{\text{KM}}$ with $\epsilon = 10^{-4}$
using the image shown in Fig.~\ref{fig:2}, we obtain the image shown
in Fig.~\ref{fig:3}. Note that the region plotted is
$20/k_{0} \times 20/k_{0}$, which is a much smaller region than that
plotted in Fig.~\ref{fig:2}. This result shows that this modified KM
method is able to image targets with subwavelength
resolution. Moreover, since the parameter $\epsilon$ is user-defined,
this high resolution is tunable.

\subsection{Range shift}

Because of the high-resolution capabilities of the modified KM method,
we are able to observe that the predicted target location is shifted
from the exact location, especially in range. The peak of the image
shown in Fig.~\ref{fig:3} is located at
$(k_{0} \hat{x},k_{0} \hat{y}) = ( 273.113, -349.770 )$ compared to
the true location at
$(k_{0} x_{0}, k_{0} y_{0}) = ( 273.713, -346.167 )$. In simulations,
we find that if the SNR is larger, the cross-range coordinate $x_{0}$
becomes exact. However, the range coordinate $y_{0}$ remains
shifted. For example, with $\text{SNR} = 13.73$ dB we obtain a
predicted target location at
$(k_{0} \hat{x}, k_{0} \hat{y}) = ( 273.713, -350.170 )$, and with
$\text{SNR} = 23.73$ dB, we obtain a predicted target location at
$(k_{0} \hat{x}, k_{0} \hat{y}) = ( 273.713, -350.170 )$.  We find
that this shift in the predicted range of the target varies with both
the size parameter, $k_{0} \alpha$, and the relative refractive index,
$n_{rel}$ used to generate the frequency dependent reflectivity.

To understand the cause of this shift in the range coordinate, we
substitute \eqref{eq:dmn} into \eqref{eq:IKM} and obtain
\begin{equation}
  I^{\text{KM}}(\ypos) = 
  \sum_{m = 1}^{M} \varrho_{0}(\omega_{m}) \sum_{n =
    1}^{N} \frac{e^{\mathrm{i} 2 \omega_{m} ( | \xpos_{n} -
      \ypos_{0}  |  - | \xpos_{n} - \ypos |)/ c}}{( 4\pi |
    \xpos_{n} - \ypos_{0} |)^{2}}.
\end{equation}
Consider a coordinate system in which the origin lies at the center of
the imaging region. The coordinates of the spatial measurements are
$\xpos_{n} = (\xi_{n}, R, H)$ for $n = 1, \dots, N$ with
$\xi_{n} = -a/2 + a (n-1)/(N-1)$.  We write
$\ypos_{0} = (x_{0}, y_{0}, 0)$ and
$\ypos = ( x_{0}, y_{0} + y, 0)$. Let $\theta$ denote the look
angle (see Fig.~\ref{fig:SAR-setup}) so that $R = L \sin\theta$ and
$H = L \cos\theta$. In the asymptotic
limit $L \to \infty$, we find that
\begin{equation}
  | \xpos_{n} - \ypos_{0} | - | \xpos_{n} - \ypos | = y \sin\theta +
  O(L^{-1}).
\end{equation}
It follows that
\begin{equation}
  I^{\text{KM}} \sim \frac{N}{(4 \pi L)^{2}} \sum_{m =
    1}^{M} \varrho_{0}(\omega_{m}) e^{\mathrm{i} 2
    \omega_{m} y \sin\theta/c},
  \label{eq:IKM-behavior}
\end{equation}
in this asymptotic limit.

Let
\begin{equation}
 a(y) = \sum_{m = 1}^{M} \varrho_{m} e^{\mathrm{i} 2 k_{m} y \sin\theta},
\end{equation}
with $\varrho_{m} = \varrho_{0}(\omega_{m})$ and $k_{m} =
\omega_{m}/c$. Using
\begin{equation}
  k_{m} = k_{0} + \frac{2 \pi B}{c} \left( - \frac{1}{2} + \frac{m -
      1}{M - 1} \right),
\end{equation}
for $m = 1, \dots, M,$ with $k_{0}$ denoting the central wavenumber,
we introduce the scaled variable $y \sin\theta = 2 \pi B Y/c$ and
consider instead
\begin{equation}
  a\left( \frac{2 \pi B}{c \sin\theta} Y \right)
  = e^{\mathrm{i} 2 \kappa_{0} Y} A(Y),
\end{equation}
where
\begin{equation}
 A(Y) =  \sum_{m = 1}^{M} \varrho_{m}
  \exp\left[ \mathrm{i} \left( - 1 + 2 \frac{m - 1}{M - 1} \right) Y
  \right],
  \label{eq:A}
\end{equation}
and $\kappa_{0} = k_{0} \sin\theta c/(2 \pi B)$.

In what follows, we make use of the following identity
\begin{equation}
  \sum_{m = r+1}^{M} \exp\left[ \mathrm{i} \left( - 1
      + 2 \frac{m - 1}{M - 1} \right) Y \right]
  = e^{\mathrm{i} r Y /(M-1)} \Psi_{r}^{M}(Y).
\end{equation}
where
\begin{equation}
  \Psi_{r}^{M}(Y) = \frac{\sin\left( \frac{M-r}{M-1} Y
    \right)}{\sin\left( \frac{Y}{M-1} \right)}.
\end{equation}
The function $\Psi_{r}^{M}(Y)$ is real and even, and it attains its
maximum of $M - r$ on $Y = 0$.  When we apply the summation by parts
formula,
\begin{equation}
  \sum_{m = 1}^{M} u_{m} v_{m} = u_{1} \sum_{m = 1}^{M} v_{m}  -
  \sum_{r = 1}^{M-1} ( u_{r+1} - u_{r} ) \sum_{m = r+1}^{M} v_{m},
\end{equation}
to \eqref{eq:A}, we find that
\begin{equation}
  A(Y) = \sum_{m = 1}^{M} \varDelta \varrho_{m} e^{\mathrm{i} (m-1)
    Y/(M-1)} \Psi_{m-1}^{M}(Y),
\end{equation}
with $\varDelta\varrho_{m} = \varrho_{m+1} - \varrho_{m}$ and
$\varrho_{0} \equiv 0$.  It follows that
\begin{multline}
  | A(Y) |^{2} = \sum_{m = 1}^{M} | \varDelta \varrho_{m} |^{2} (
  \Psi_{m-1}^{M}(Y) )^{2} + 2 \sum_{m = 1}^{M-1} \sum_{r = m + 1}^{M}
  \bigg\{ \text{Re}\left[ \varDelta \varrho_{m}^{\ast} \varDelta
    \varrho_{r}
  \right] \cos\left( \frac{r - m}{M - 1} Y \right)  \Psi_{m-1}^{M}(Y)
  \Psi_{r-1}^{M}(Y) \bigg\}\\
  - 2 \sum_{m = 1}^{M-1} \sum_{r = m + 1}^{M} \bigg\{ \text{Im}\left[
    \varDelta \varrho_{m}^{\ast} \varDelta \varrho_{r} \right]
  \sin\left( \frac{r - m}{M-1} Y \right) \Psi_{m-1}^{M}(Y)
  \Psi_{r-1}^{M}(Y) \bigg\}.
  \label{eq:F-squared}
\end{multline}
Note that in \eqref{eq:F-squared} the first two sums are even
functions of $Y$ and the third sum is odd in $Y$. That third sum plays
a key role in the range shift.

To compute an estimate for where $|A(Y)|^{2}$ attains its maximum, we
expand the functions of $Y$ in each of the three sums in
\eqref{eq:F-squared} about $Y = 0$ and keep terms up to
$O(Y^{2})$. Combining these results yields a quadratic approximation
for $|A(Y)|^{2}$. By computing the critical point of this quadratic
approximation, we find that this approximation attains its maximum on
$\hat{Y} = -3 (M-1) \alpha_{1}/\alpha_{2}$ where
\begin{equation}
  \alpha_{1} = \sum_{m = 1}^{M-1} \sum_{r = m+1}^{M} \text{Im}[
  \varDelta\varrho_{m} \varDelta\varrho_{r} ] ( r - m )
  (M - m + 1) (M - r + 1),
\end{equation}
and
\begin{multline}
  \alpha_{2} = \sum_{m = 1}^{M} | \varDelta\varrho_{m} |^{2} \left[ (
    M - m + 1 )^{4} - (M - m + 1)^{2} \right]\\
  + \sum_{m = 1}^{M-1} \sum_{r = m+1}^{M} \text{Re}[
  \varDelta\varrho_{m}^{\ast} \varDelta\varrho_{r} ] \left( M^{2} + 2
    M - 3 m r 
  + m ( 2m - M - 1 ) + r ( 2r - M - 1) \right) .
\end{multline}

\begin{figure}[htb]
  \centering
  \includegraphics[width=0.48\linewidth]{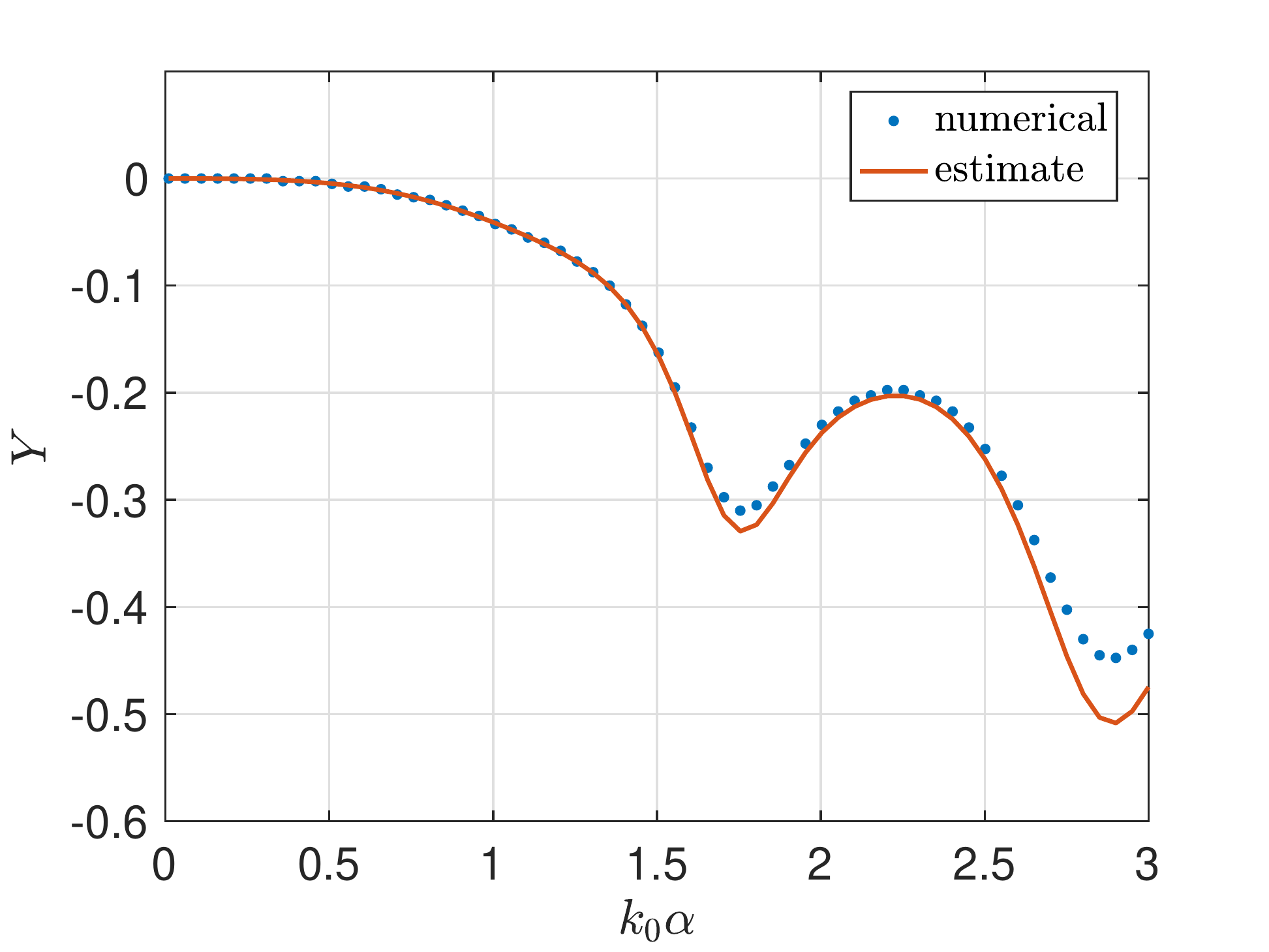}
  \caption{A comparison of the range coordinate of the target
    predicted using KM (numerical), and by computing the estimate
    $\hat{Y}$ (estimate) over reflectivities computed using spheres
    with $n_{rel} = 1.4$ over a range of different sizes,
    $k_{0}\alpha$. }
  \label{fig:4}
\end{figure}

This critical point $\hat{Y}$ gives an estimate for the range shift of
the target location predicted by KM.  We show a comparison of the
numerically determined location of the predicted target and this
estimate in Fig.~\ref{fig:4}. This comparison is done using the
frequency dependent reflectivity of a sphere with relative refractive
index $1.4$ for various non-dimensional sizes, $k_{0} \alpha$. 

\begin{figure}[htb]
  \centering
  \includegraphics[width=0.48\linewidth]{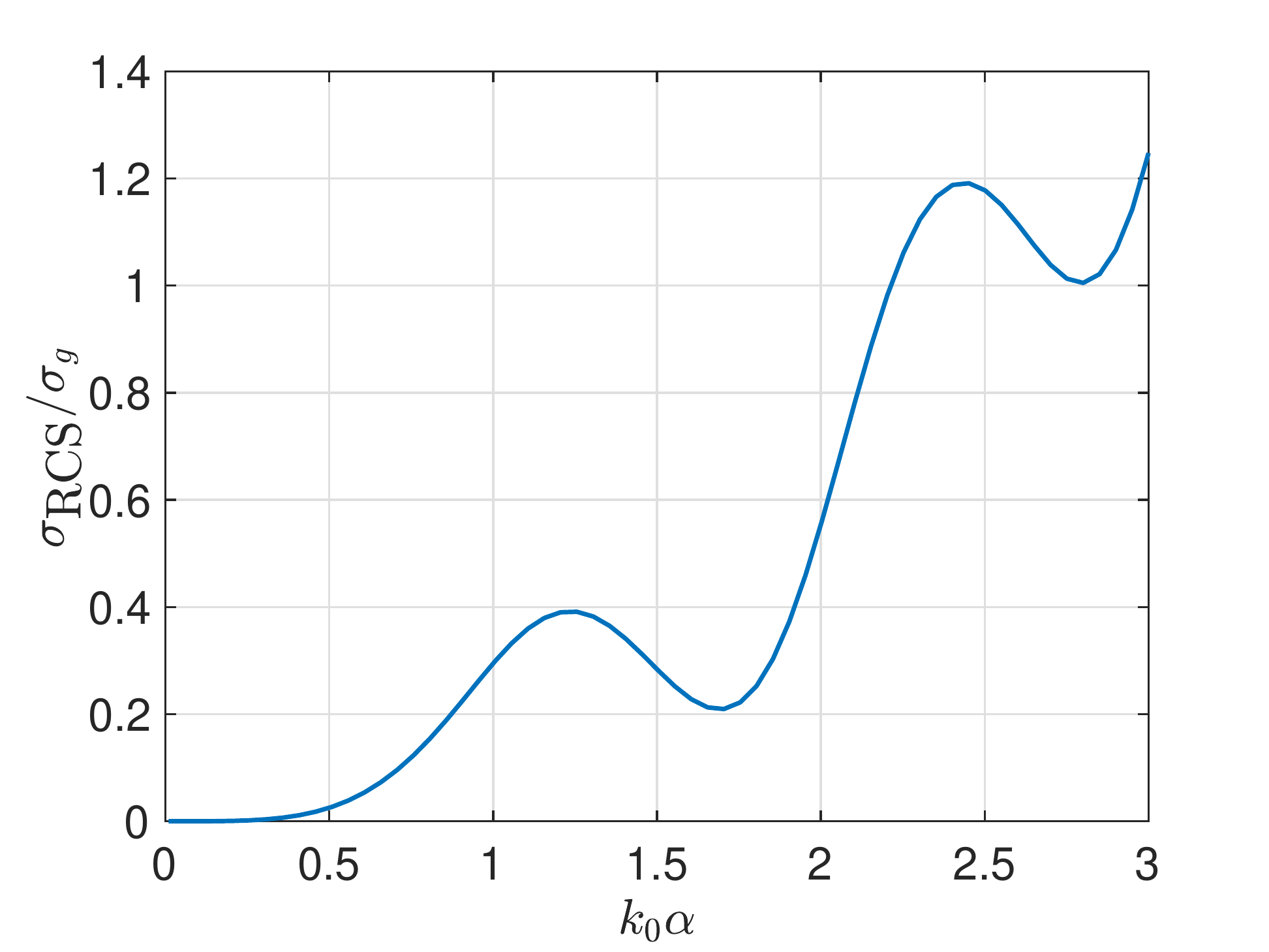}
  \caption{The RCS $\sigma_{\text{RCS}}$ normalized by the geometric
    cross-section $\sigma_{g} = \pi \alpha^{2}$ evaluated on the
    central frequency $\omega_{0}$ for spheres with $n_{rel} = 1.4$
    over a range of sphere sizes $k_{0} \alpha$.}
  \label{fig:5}
\end{figure}

These results show that the estimate accurately captures the behavior
of this range shift over a broad range of sphere sizes. However, the
error of this estimate grows with the sphere size, but especially
where the range shift oscillates. These oscillations are presumably
due to the complex scattering behavior of large spheres
($k_{0} \alpha > 1$) that exhibit phenomena such as Mie resonances. In
Fig.~\ref{fig:5} we show the RCS evaluated on the central frequency
$\omega_{0}$ normalized by the geometric cross-section,
$\sigma_{g} = \pi \alpha^{2}$ for spheres with $n_{rel} = 1.4$ over
the same range of $k_{0} \alpha$ plotted in Fig.~\ref{fig:4}. Note
that the behavior of the range shifts shown in Fig.~\ref{fig:4}
closely follow the behavior of the RCS shown in Fig.~\ref{fig:5}. In
this way, we see that the range shift in the predicted range of the
target by KM is inherently connected to fundamental scattering
properties of the target.

\subsection{Radar cross-section}

We now seek to recover $\varrho(\omega_{m})$ for $m = 1, \dots,
M$. Suppose we have evaluated \eqref{eq:IKM} and produced an image
that identifies a target and its predicted location,
$\hat{\ypos}_{0}$. We evaluate
\begin{equation}
  \phi_{m} = \frac{1}{N} \sum_{n = 1}^{N} d_{mn} (4 \pi | \xpos_{n} -
  \hat{\ypos}_{0} |)^{2} e^{-\mathrm{i} 2 \omega_{m} |
    \xpos_{n} - \hat{\ypos}_{0} |/c},
  \label{eq:phi-m}
\end{equation}
for $m = 1, \dots, M$. Substituting \eqref{eq:dmn} into
\eqref{eq:phi-m} yields
\begin{equation}
  \phi_{m} = \varrho_{0}(\omega_{m}) \frac{1}{N} \sum_{n = 1}^{N}
  \frac{| \xpos_{n} - \hat{\ypos}_{0} |^{2}}{| \xpos_{n} - \ypos_{0}
    |^{2}} e^{\mathrm{i} 2 \omega_{m} \varDelta \tau_{n}},
\end{equation}
with
$\varDelta\tau_{n} = ( | \xpos_{n} - \ypos_{0} | - | \xpos_{n} -
\hat{\ypos}_{0} | )/ c$.

In the asymptotic limit $L \to \infty$, we find using the same
expansions used above that
\begin{equation}
  \phi_{m} \sim \varrho_{0}(\omega_{m})
  e^{\mathrm{i} 2 \omega_{m} \varDelta y \sin\theta/ c},
\end{equation}
for $m = 1, \dots, M$. Here, $\varDelta y$ denotes the range shift
associated with the predicted location $\hat{\ypos}_{0}$. Because the
range shift $\varDelta y$ is not known, we cannot remove the factor of
$e^{\mathrm{i} 2 \omega_{m} \varDelta y \sin\theta / c}$ from the
expression above. Therefore, we cannot recover the complex values of
$\varrho_{0}(\omega_{m})$. However, we find that
\begin{equation}
  | \phi_{m} |^{2} \sim | \varrho_{0}(\omega_{m}) |^{2}.
\end{equation}
Therefore, we recover the RCS given the predicted location of the
target by KM through evaluation of
\begin{equation}
  \hat{\sigma}_{\text{RCS}}(\omega_{m}) = 4 \pi | \phi_{m} |^{2}.
  \label{eq:RCS-estimate}
\end{equation}

\begin{figure}[htb]
  \centering
  \includegraphics[width=0.48\linewidth]{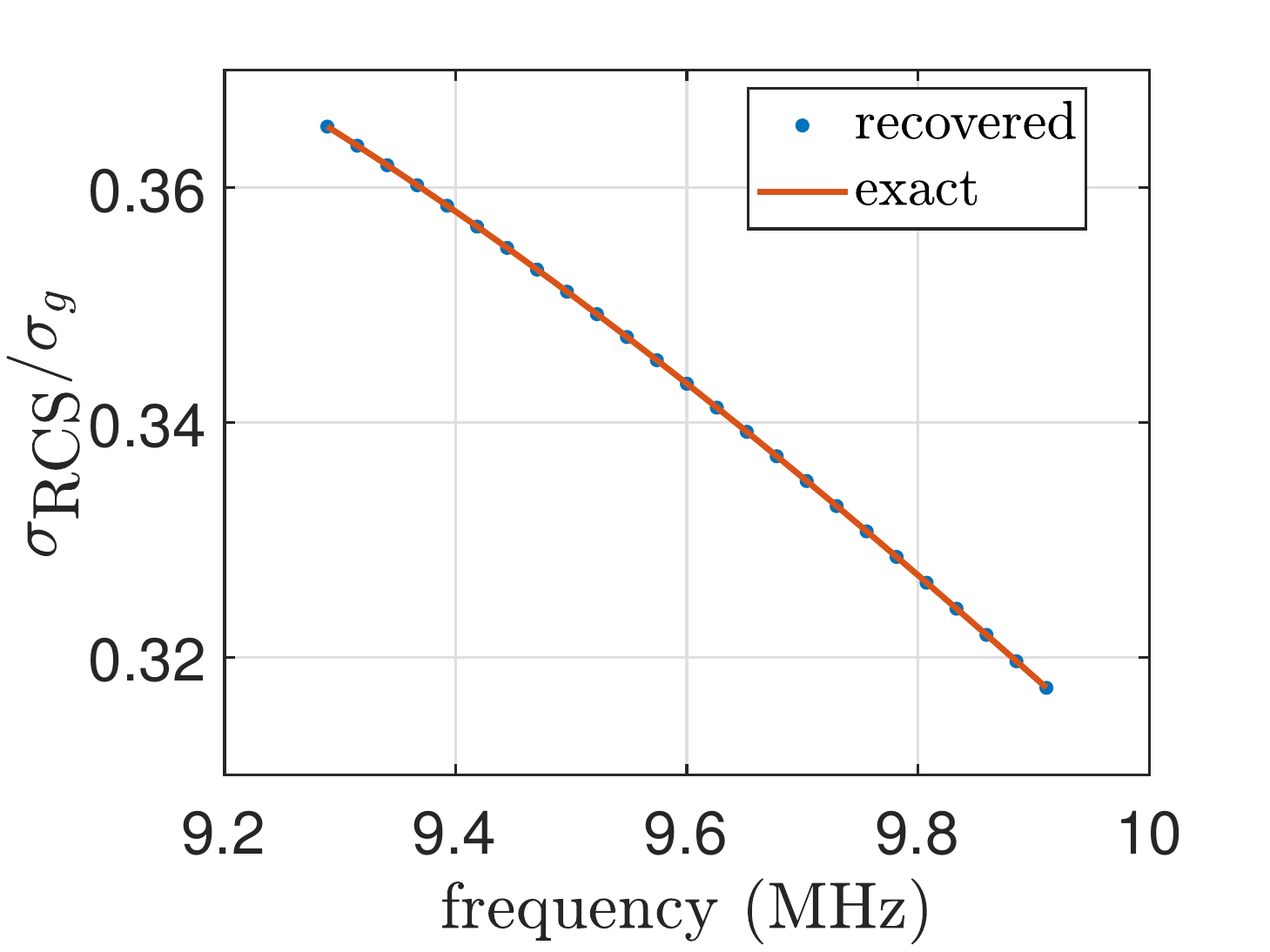}
  \caption{The recovered RCS using \eqref{eq:RCS-estimate} normalized
    by the geometric cross-section $\sigma_{g} = \pi \alpha^{2}$ for
    the reflectivity used in Figs.~\ref{fig:2} and \ref{fig:3}
    corresponding to $k_{0} \alpha = 1.4$ and $n_{rel} = 1.4$. To
    compute $\hat{\sigma}_{\text{RCS}}(\omega)$, we estimate
    $\hat{\ypos}_{0}$ by finding where $I_{\epsilon}^{\text{KM}}$
    attains its maximum value on the mesh used for plotting.}
  \label{fig:6}
\end{figure}

In Fig.~\ref{fig:6} we show the estimated RCS using
\eqref{eq:RCS-estimate} for the same data used in Figs.~\ref{fig:2}
and \ref{fig:3}. To estimate $\hat{\ypos}_{0}$ we use the mesh
location used to plot those images where $I_{\epsilon}^{\text{KM}}$
attains its maximum value which is
$(\hat{x},\hat{y}) = ( 1.357, -1.738 )$ cm compared to the true
location $(x_{0}, y_{0}) = ( 1.360, -1.720 )$ cm. The exact RCS
computed from the reflectivity is plotted for comparison. The
estimated RCS is indistinguishable from the exact RCS and the relative
error is on the order of $10^{-4}$.

\begin{figure}[htb]
  \centering
  \includegraphics[width=0.48\linewidth]{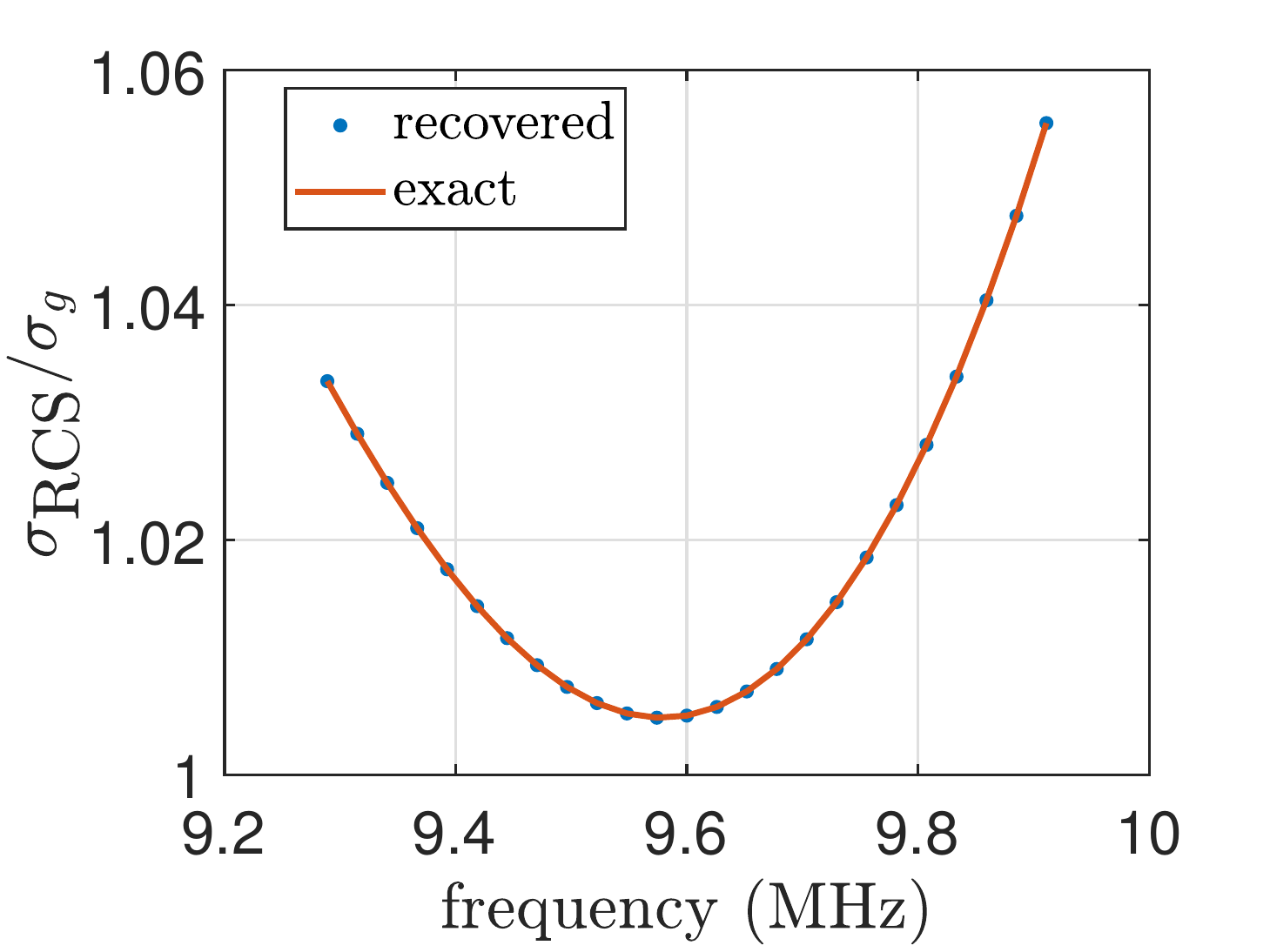}
  \caption{The recovered RCS using \eqref{eq:RCS-estimate} normalized
    by the geometric cross-section $\sigma_{g} = \pi \alpha^{2}$ for
    the reflectivity 
    corresponding to $k_{0} \alpha = 2.8$ and $n_{rel} = 1.4$. To
    compute $\hat{\sigma}_{\text{RCS}}(\omega)$, we estimate
    $\hat{\ypos}_{0}$ by finding where $I_{\epsilon}^{\text{KM}}$
    attains its maximum value on the mesh used for plotting.}
  \label{fig:7}
\end{figure}

We consider a target whose reflectivity is computed for a sphere of
size $k_{0} \alpha = 2.8$ and relative refractive index
$n_{rel} = 1.4$. Measurement noise was added so that
$\text{SNR} = 3.72$ dB. The location predicted by finding the mesh
point on which $I_{\epsilon}^{\text{KM}}$ attains its maximum is
$(\hat{x},\hat{y}) = ( 1.358, -1.797 )$ cm. We estimate the RCS
through evaluation of \eqref{eq:RCS-estimate} using this predicted
target location. The results are plotted in Fig.~\ref{fig:7}. Note
that the RCS for this problem is markedly different from that shown in
Fig.~\ref{fig:6}. Nonetheless, the estimated RCS is still accurate
with a relative error on the order of $10^{-5}$.

\section{Multiple targets}
\label{sec:multiple-targets}

Suppose now the imaging region contains $Q$ dispersive point targets
at locations $\ypos_{q}$ with reflectivities $\varrho_{q}(\omega)$ for
$q = 1, \dots, Q$. Assuming that these targets scatter independently,
measurements are modeled according to
\begin{equation}
  d_{mn} = \sum_{q = 1}^{Q} \varrho_{q}(\omega_{m}) \frac{e^{\mathrm{i} 2
      \omega_{m} | \xpos_{n} - \ypos_{q} |/c}}{(4 \pi | \xpos_{n} -
    \ypos_{q} |)^{2}}.
  \label{eq:multiple-targets}
\end{equation}
From our results for a single target, we anticipate that evaluating
the KM imaging function given in \eqref{eq:IKM} will identify and
locate targets under the condition that these targets are not too
close to one another as measured with respect to the resolution
produced by KM for a single target.

\begin{figure}[htb]
  \centering
  \includegraphics[width=0.48\linewidth]{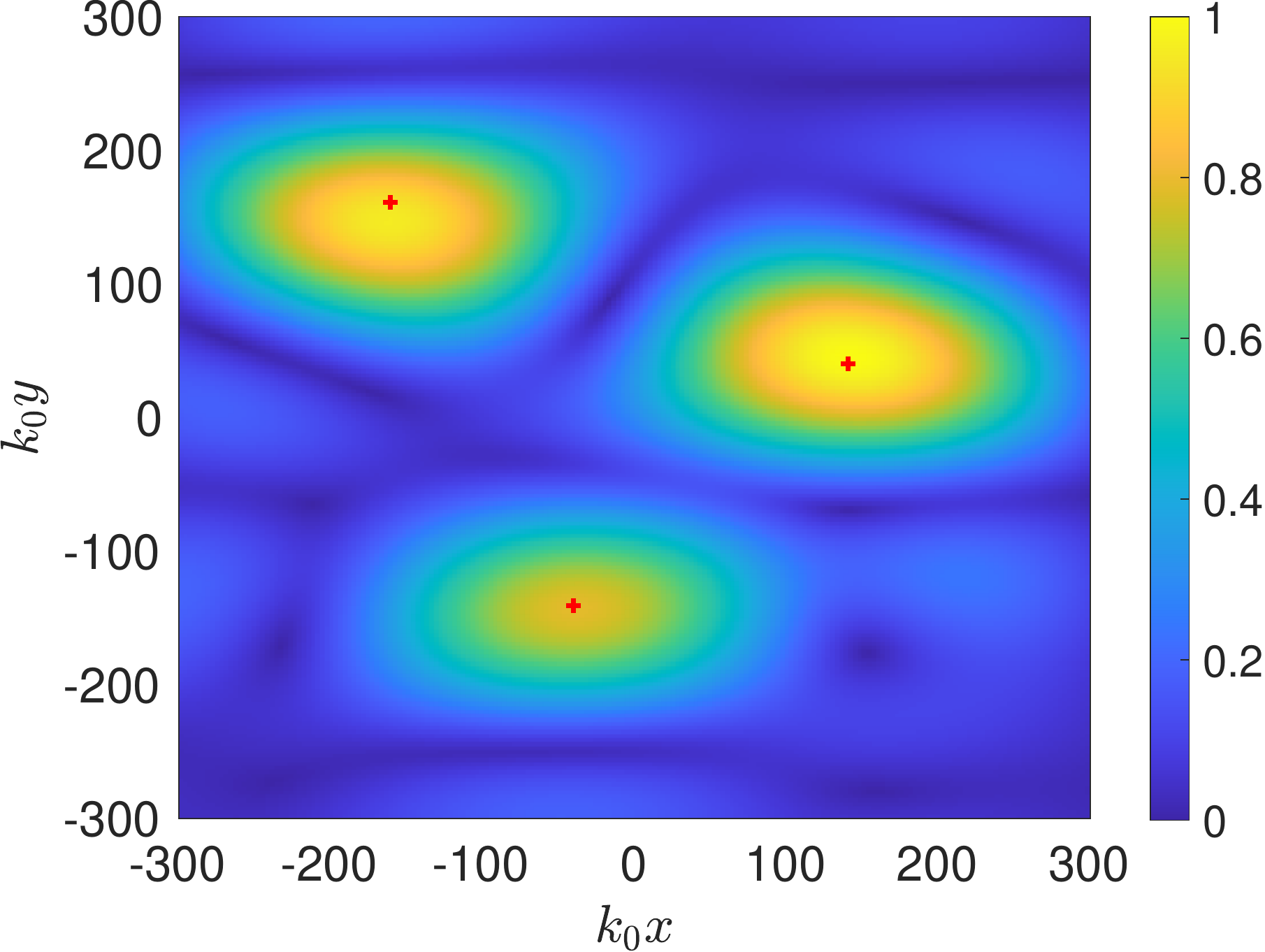}
  \caption{Imaged produced by KM through evaluation of \eqref{eq:IKM}
    over an imaging region containing 3 dispersive point targets whose
    exact locations are plotted as red ``+'' symbols.}
  \label{fig:8}
\end{figure}

In Fig.~\ref{fig:8} we show a result of evaluating \eqref{eq:IKM} over
an imaging region containing three different dispersive point
targets. The first target is located at
$( k_{0} x_{1}, k_{0} y_{1}, k_{0} z_{1} ) = ( 140.882, 40.252, 0
)$. Its reflectivity $\varrho_{1}(\omega)$ is computed using a sphere
of size $k_{0} \alpha_{1} = 0.8$ and relative refractive index
$n_{rel,1} = 1.8$. The second target is located at
$( k_{0} x_{2}, k_{0} y_{2}, k_{0} z_{2} ) = ( -40.252, -140.882, 0
)$. Its reflectivity $\varrho_{2}(\omega)$ is computed using a sphere
of size $k_{0} \alpha_{1} = 1.2$ and relative refractive index
$n_{rel,2} = 1.4$. The third target is located at
$( k_{0} x_{3}, k_{0} y_{3}, k_{0} z_{3} ) = ( -161.008, 161.008, 0
)$. Its reflectivity $\varrho_{3}(\omega)$ is computed using a sphere
of size $k_{0} \alpha_{3} = 1.8$ and relative refractive index
$n_{rel,3} = 1.4$. Measurement noise was added to the data so that
$\text{SNR} = 22.84$ dB.  Figure~\ref{fig:8} shows three distinct
peaks in the vicinity of the three targets whose locations are plotted
as red ``+'' symbols.

\begin{figure}[htb]
  \centering
  \includegraphics[width=0.32\linewidth]{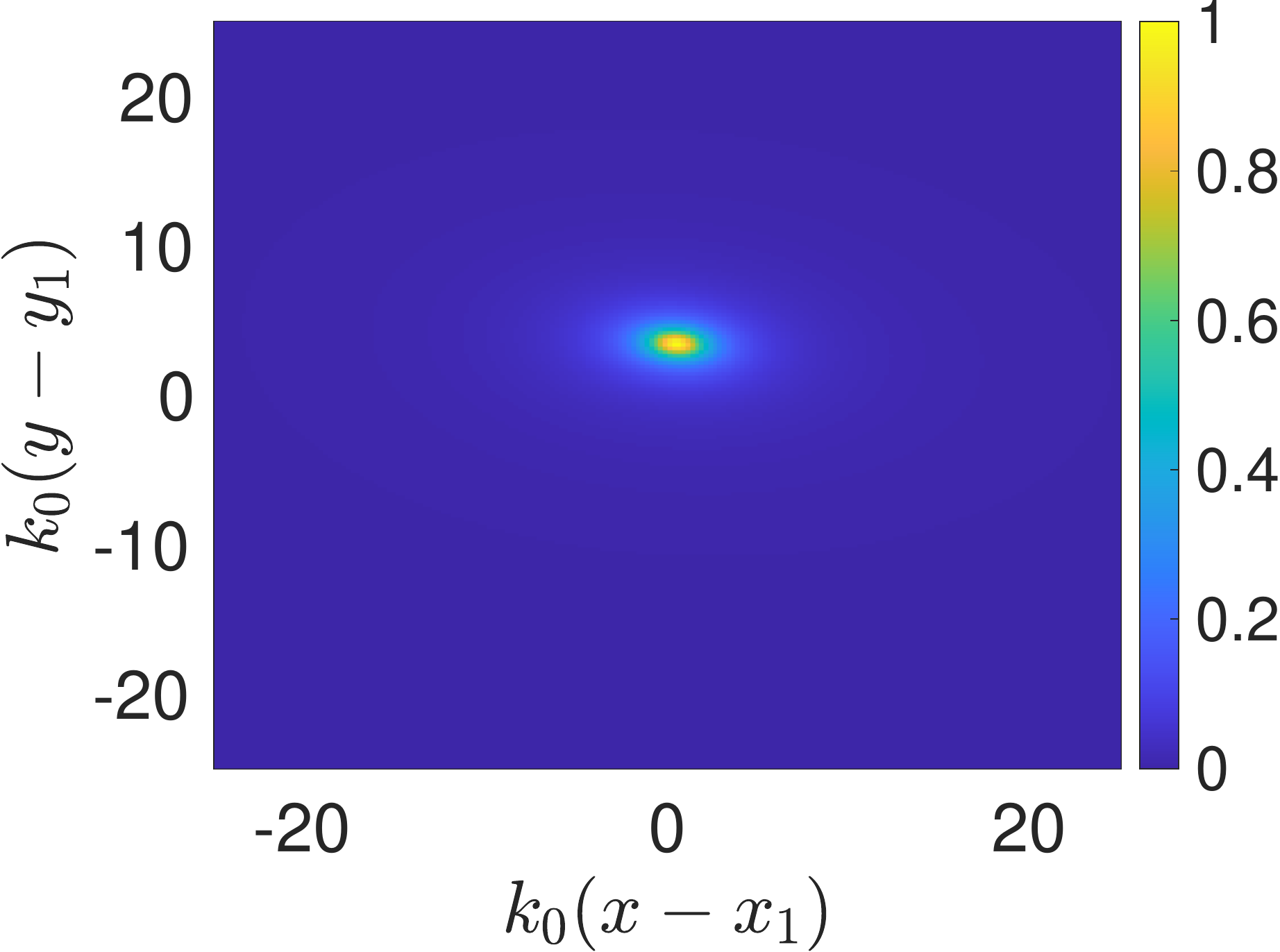}
  \includegraphics[width=0.32\linewidth]{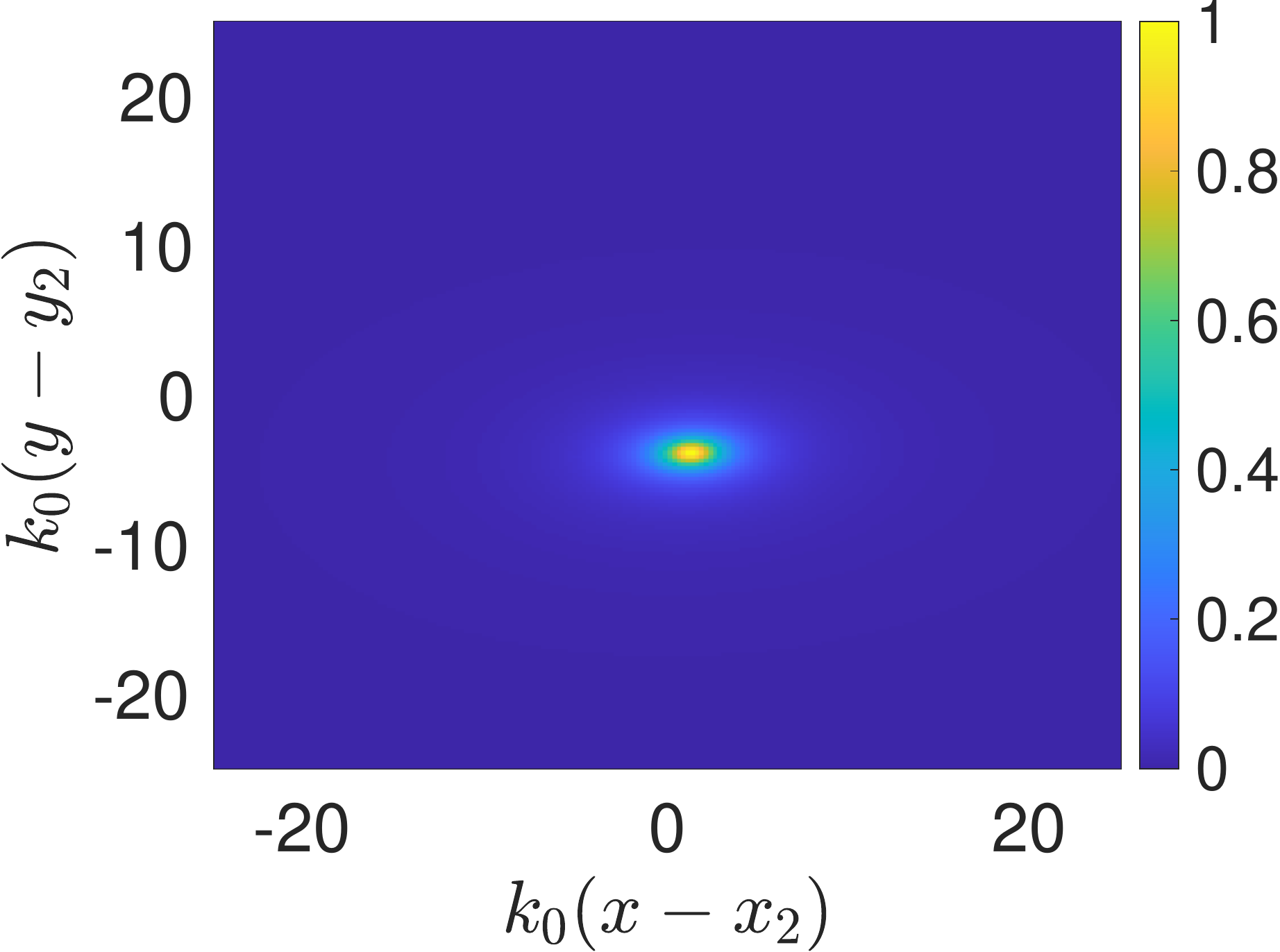}
  \includegraphics[width=0.32\linewidth]{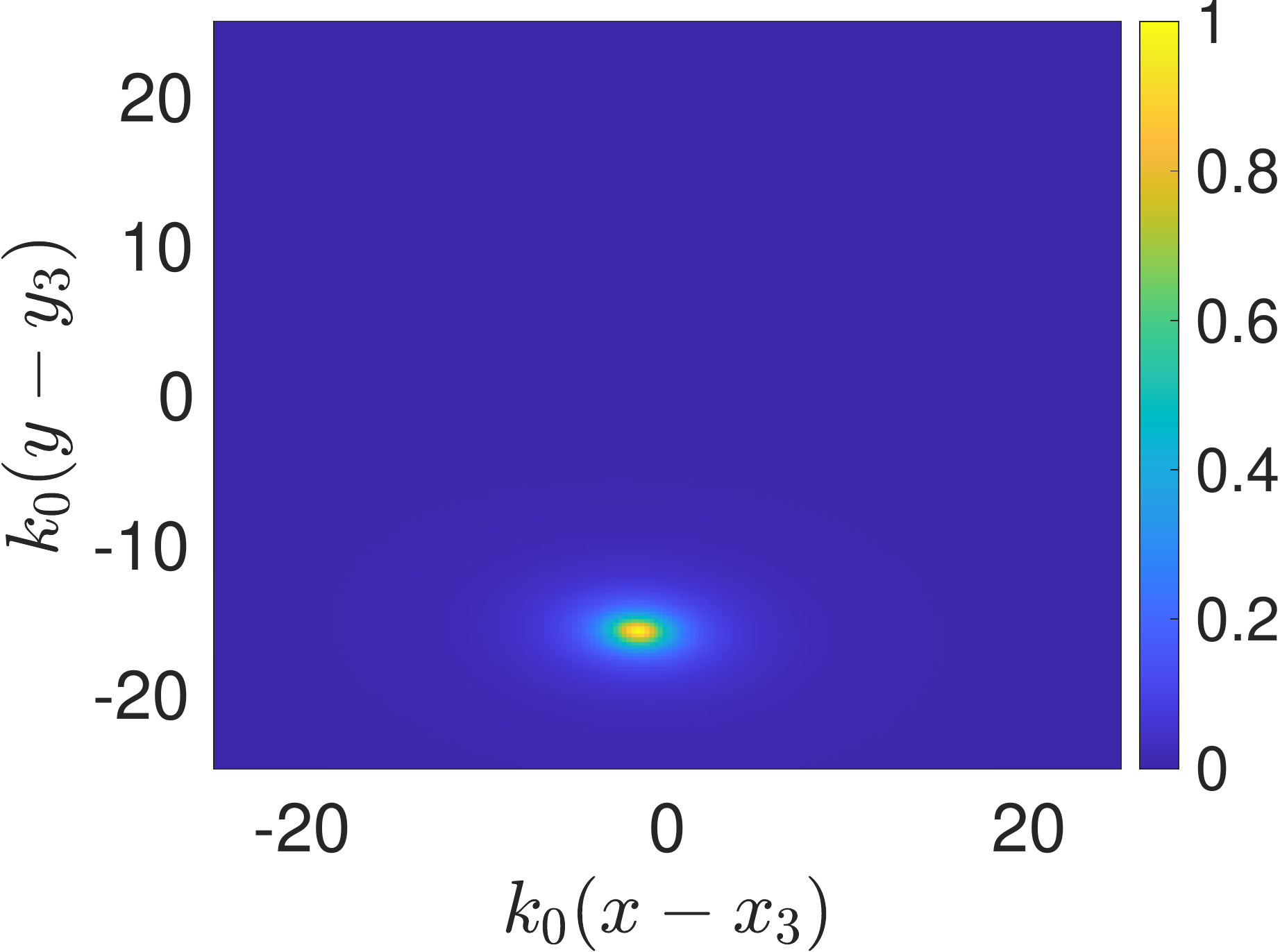}
  \caption{Images produced using modified KM given in \eqref{eq:rKM}
    in $50/k_{0} \times 50/k_{0}$ sub-regions centered about each of
    the exact target locations. The left plot corresponds to target
    1, the center plot corresponds to target 2 and the right plot
    corresponds to target 3.}
  \label{fig:9}
\end{figure}

To obtain high-resolution images of individual targets, we consider
$50/k_{0} \times 50/k_{0}$ sized sub-regions about each of the peaks
shown in Fig.~\ref{fig:8}. We normalize the portion of the image
contained in each of those sub-regions so that the maximum value
contained in that sub-region is unity. Then we apply \eqref{eq:rKM}
with $\epsilon = 10^{-4}$. Those results appear in
Fig.~\ref{fig:9}. 

The modified KM images produce high resolution images of the
targets. However, we observed that the predicted target locations are
shifted in both range and cross-range from the exact target
locations. The predicted location for target 1 is
$(k_{0} \hat{x}_{1}, k_{0} \hat{y}_{1}) = ( 141.382, 43.502 )$, for
target 2 is
$(k_{0} \hat{x}_{2}, k_{0} \hat{y}_{2}) = ( -39.002, -144.882 )$, and
for target 3 is
$(k_{0} \hat{x}_{3}, k_{0} \hat{y}_{3}) = ( -162.758, 145.008 )$. These
shifts in the predictions from the exact locations are approximately
$2 \lambda_{0}$.

To understand why these shifts in predicted target locations occur in
both range and cross-range, suppose we evaluate \eqref{eq:IKM} on
$\ypos_{1}$, the exact location for target one. The result is
\begin{equation}
  I^{\text{KM}}(\ypos_{1}) = \sum_{m = 1}^{M} \sum_{n = 1}^{N} \left[
    \frac{\varrho_{1}(\omega_{m})}{( 4 \pi | \xpos_{n} - \ypos_{1} |
      )^{2}} + \sum_{q = 2}^{Q} \varrho_{q}(\omega_{m}) \frac{|
      \xpos_{n} - \ypos_{1} |^{2}}{| \xpos_{n} - \ypos_{q} |^{2}}
    e^{\mathrm{i} 2 \omega_{m} \varDelta\tau_{1q}^{n}} \right],
\end{equation}
where
$\varDelta\tau_{pq}^{n} = ( | \xpos_{n} - \ypos_{q} | - | \xpos_{n} -
\ypos_{p} | )/c$. From this result we see that in addition to the
contribution made by target 1, we obtain small, but non-trivial
contributions by all other targets, each of which carries a phase
associated with $\varDelta\tau_{q,p}^{n}$. Therefore, upon computing
$|I^{\text{KM}}|$, those phases mix leading to shifts in the predicted
target positions.

Even though the predicted locations of targets are not exact, we seek
to recover the RCS of each of the targets. Let $\hat{\ypos}_{p}$
denote the approximate location of target $p$. Let
\begin{equation}
  \phi_{m}(\hat{\ypos}_{p}) = \frac{1}{N} \sum_{n = 1}^{N} d_{mn} ( 4
  \pi | \xpos_{n} - \hat{\ypos}_{p} | )^{2} e^{-\mathrm{i} 2
    \omega_{m} | \xpos_{n} - \hat{\ypos}_{p} | / c }.
\end{equation}
Substituting \eqref{eq:multiple-targets} into this expression, we
obtain
\begin{equation}
  \phi_{m}(\hat{\ypos}_{p}) = \sum_{q = 1}^{Q} a_{pq}(\omega_{m})
  \varrho_{q}(\omega_{m}),
  \label{eq:linear-system}
\end{equation}
where
\begin{equation}
  a_{pq}(\omega_{m}) = \frac{1}{N} \sum_{n = 1}^{N} \frac{| \xpos_{n}
    - \hat{\ypos}_{p}  |^{2}}{| \xpos_{n} - \ypos_{q} |^{2}}
  e^{\mathrm{i} 2 \omega_{m}  \varDelta\hat{\tau}_{q,p}^{n}},
\end{equation}
and
$\varDelta\hat{\tau}_{p q}^{n} = ( | \xpos_{n} - \ypos_{q} | - |
\xpos_{n} - \hat{\ypos}_{p} | )/c$. Equation \eqref{eq:linear-system}
is a linear system for the unknown reflectivities
$\varrho_{q}(\omega_{m})$. Although $a_{pq}(\omega_{m})$ uses the
predicted target position $\hat{\ypos}_{p}$, it uses the exact target
positions $\ypos_{q}$. Since we do not have access to those exact
target locations, we instead consider the linear system,
\begin{equation}
  \phi_{m}(\hat{\ypos}_{p}) = \sum_{q = 1}^{Q}
  \tilde{a}_{pq}(\omega_{m}) \tilde{\varrho}_{q}(\omega_{m}),
  \label{eq:approx-linear-system}
\end{equation}
where
\begin{equation}
  \tilde{a}_{pq}(\omega_{m}) = \frac{1}{N} \sum_{n = 1}^{N} \frac{| \xpos_{n}
    - \hat{\ypos}_{p}  |^{2}}{| \xpos_{n} - \hat{\ypos}_{q} |^{2}}
  e^{\mathrm{i} 2 \omega_{m}  \varDelta\tilde{\tau}_{q,p}^{n}},
\end{equation}
with
$\varDelta\tilde{\tau}_{p q}^{n} = ( | \xpos_{n} - \hat{\ypos}_{q} | -
| \xpos_{n} - \hat{\ypos}_{p} | )/c$, and
\begin{equation}
  \tilde{\varrho}_{q}(\omega_{m}) = \varrho_{q}(\omega_{m}) \frac{| \xpos_{n}
    - \hat{\ypos}_{q}  |^{2}}{| \xpos_{n} - \ypos_{q} |^{2}}
  e^{\mathrm{i} \omega_{m} ( \varDelta\tau_{pq}^{n} -
    \varDelta\hat{\tau}_{pq}^{n} )}.
\end{equation}
When the predicted target locations are close, we expect that
$| \xpos_{n} - \hat{\ypos}_{q} |^{2}/| \xpos_{n} - \ypos_{q} |^{2}
\approx 1$. The difference in phase,
$\varDelta\tau_{pq}^{n} - \varDelta\hat{\tau}_{pq}^{n}$, may be
significant, so we expect that we will not be able to recover
$\varrho_{q}(\omega_{m})$ from
$\tilde{\varrho}_{q}(\omega_{m})$. However, in the asymptotic limit as
$L \to \infty$, we find that the RCS for the $q$th target is
\begin{equation}
  \sigma_{\text{RCS}, q}(\omega_{m}) = 4 \pi |
  \tilde{\varrho}_{q}(\omega_{m}) |^{2} + O(L^{-1}).
  \label{eq:RCS-multiple-targets}
\end{equation}
Hence, we use this leading behavior to estimate the RCS of the
targets.

To summarize, we give the following procedure for estimating the RCS
for each of the targets identified in the imaging region.
\begin{enumerate}

\item Evaluate \eqref{eq:IKM} over the imaging region to identify
  targets.

\item Evaluate \eqref{eq:rKM} in sub-regions to estimate the locations
  of individual targets.

\item Solve the linear system \eqref{eq:approx-linear-system} and
  obtain $\tilde{\varrho}_{q}(\omega_{m})$ for $q = 1, \dots, Q$ and
  $m = 1, \dots, M$.

\item Evaluate \eqref{eq:RCS-multiple-targets} to obtain estimates for
  the RCS for each of the $Q$ targets.

\end{enumerate}

\begin{figure}[htb]
  \centering
  \includegraphics[width=0.99\linewidth]{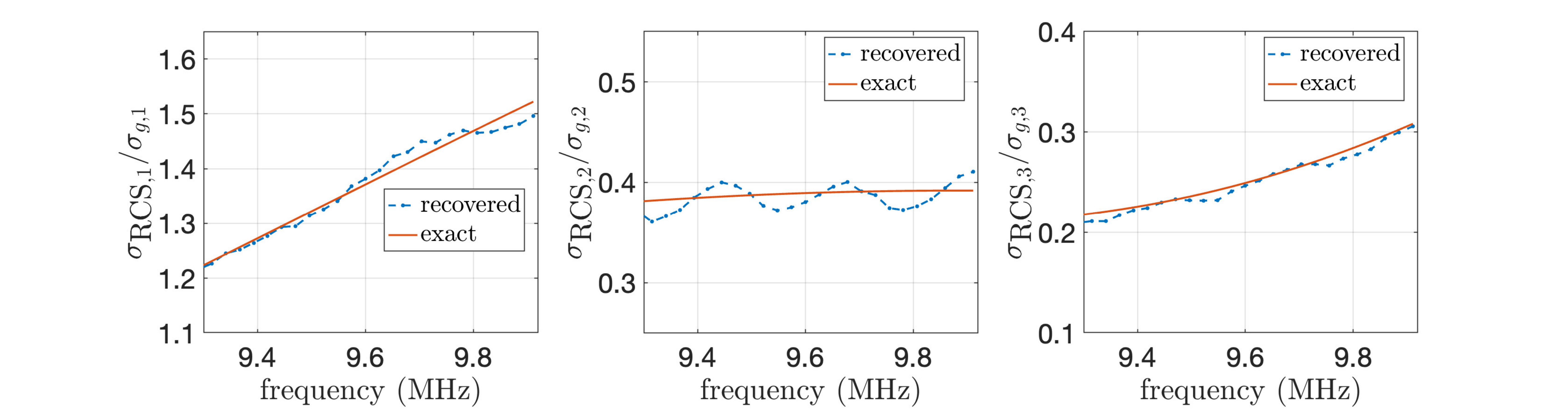}
  \caption{The recovered RCS for the three targets normalized by the
    corresponding geometric cross-section for the three targets whose
    location is recovered from Fig \ref{fig:7}. Data with
    $\text{SNR} = 22.84$ dB. The left plot corresponds to target 1,
    the center plot corresponds to target 2 and the right plot
    corresponds to target 3.}
  \label{fig:10}
\end{figure}

\begin{figure}[htb]
  \centering
  \includegraphics[width=0.99\linewidth]{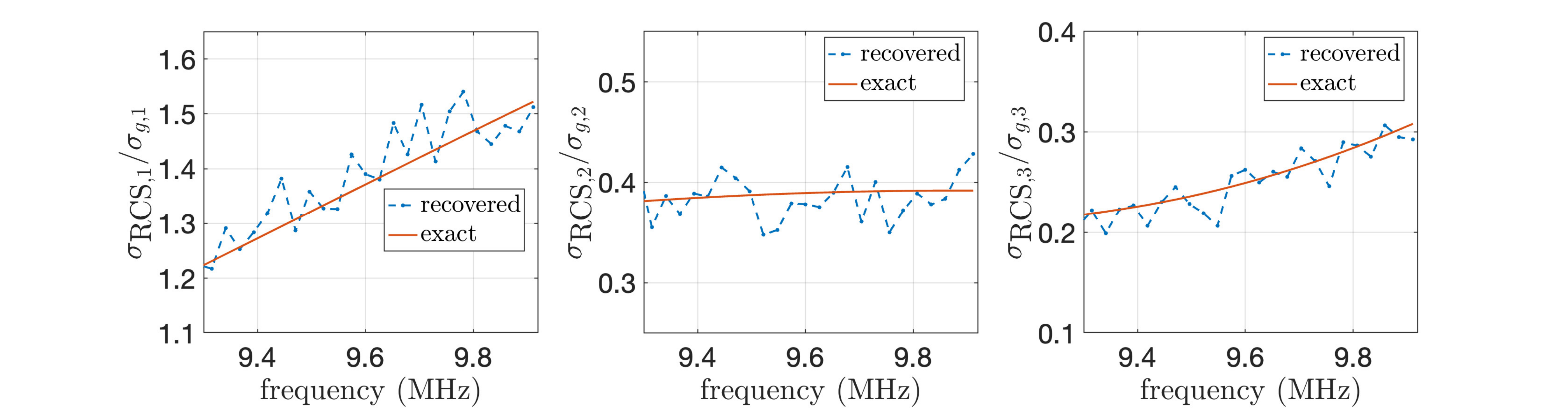}
  \caption{The recovered RCS for the three targets normalized by the
    corresponding geometric cross-section. Data with
    $\text{SNR} = 12.84$ dB. The left plot corresponds to target 1,
    the center plot corresponds to target 2 and the right plot
    corresponds to target 3.}
  \label{fig:11}
\end{figure}

The results for the recovered RCS for the three targets shown in
Figs.~\ref{fig:8} and \ref{fig:9} are shown in
Fig.~\ref{fig:10}. These results are more noisy than the ones obtained
for the single target case but their accuracy is still sufficient to
help us characterize targets of different materials/size.  As the SNR
of data decreases, the locations of the targets are recovered with the
same precision but the recovered RCS's are more noisy. For data with
$\text{SNR} = 12.84$ dB, the predicted location for target 1 is
$(k_{0} \hat{x}_{1}, k_{0} \hat{y}_{1}) = ( 141.632, 43.002 )$, for
target 2 is
$(k_{0} \hat{x}_{2}, k_{0} \hat{y}_{2}) = ( -39.502, -144.632 )$, and
for target 3 is
$(k_{0} \hat{x}_{3}, k_{0} \hat{y}_{3}) = ( -163.508, 145.508 )$ and
the recovered RCS are shown in Figure \ref{fig:11}.

\begin{figure}[htb]
  \centering
  \includegraphics[width=0.99\linewidth]{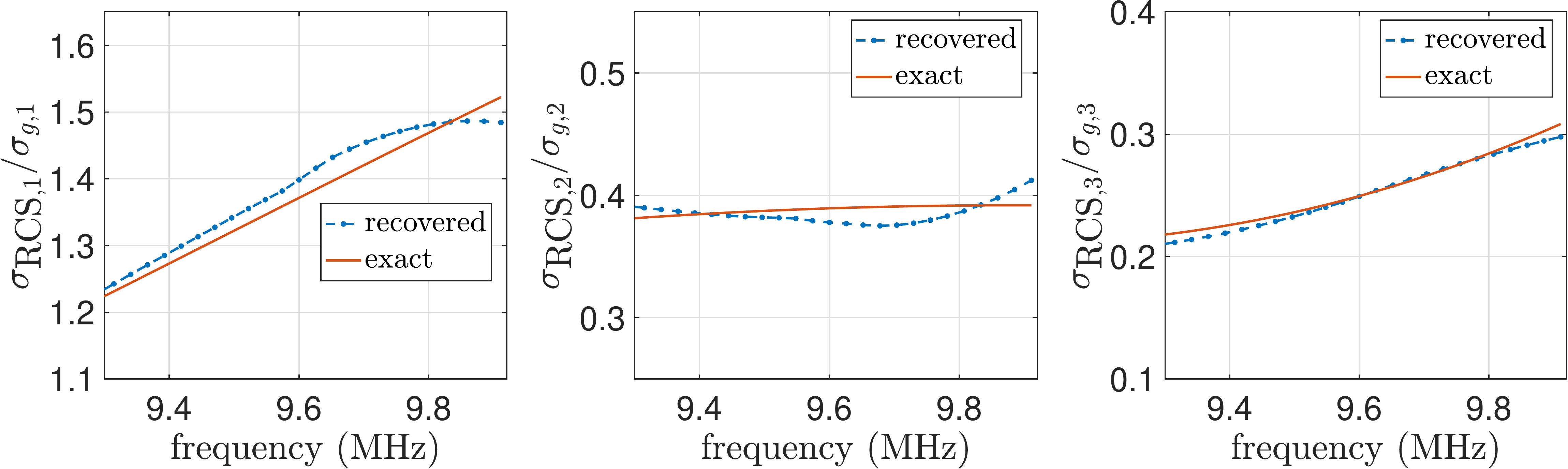}
  \caption{The recovered RCS for the three targets normalized by the
    corresponding geometric cross-section. Data with
    $\text{SNR} = 12.84$ dB. Smoothing using quadratic regression over
    the results shown in Fig \ref{fig:9}. The left plot corresponds
    for target 1, the center plot corresponds to target 2 and the
    right plot corresponds to target 3.}
  \label{fig:12}
\end{figure}

We observe in Fig.~\ref{fig:11} that the recovered RCS oscillates due
to the measurement noise. A smoothing estimate can be obtained using
quadratic regression as illustrated by the results in
Fig.~\ref{fig:12}. Those smoothed results effectively capture
the behaviors of the RCS for the individual targets.

\section{Conclusions}
\label{sec:conclusions}

We have introduced a dispersive point target model in which the
reflectivity is dependent on frequency. We have used scalar wave
scattering by a dielectric sphere to model targets of different sizes
and materials. Then we have applied KM and our recent modification to
KM to obtain high-resolution images of dispersive point targets from
SAR measurements. For a single dispersive point target, we observe a
shift in the predicted location of the target in the range
coordinate. We have computed an accurate estimate for this range shift
which is in terms of the frequency dependent reflectivity. Despite the
range shift in the predicted location of a single dispersive point
target, we are able to recover its radar cross-section (RCS) using
that prediction.

When we apply KM and its modification to an imaging region containing
multiple dispersive point targets, we find that we can image each of
the target locations provided that they are far enough apart from one
another with respect to the resolution of KM. We have shown that our
modification to KM works in sub-regions that isolate an individual
target. Those high resolution images of individual targets reveal
that the predicted locations of the targets are shifted in range and
cross-range. To recover the RCS for each of the targets, we introduce
a linear system using the predicted locations. Our numerical results
show that the recovery of RCS's for multiple targets is much more
sensitive to noise. By applying smoothing to those RCS results, we
obtain good approximations that allow one to distinguish qualitative
differences between the different targets.

By introducing the dispersive point target model and developing
methods for imaging dispersive point targets, we have opened
opportunities for target classification. 
Indeed, by recovering the RCS
as a function of frequency, we may be able to distinguish targets with
different characteristics such as sizes or material properties. We
believe that this opportunity to classify in addition 
to target detection and locatization is useful for a broad variety of SAR imaging
applications.

\appendix

\section{Scalar wave scattering by a sphere}

\label{sec:appendix}

We briefly describe scalar wave scattering by a sphere and explain how
we generated different frequency dependent reflectivities from this
problem. For a fixed frequency, let $k_{0}$ denote the wavenumber in
the exterior to a sphere of radius $\alpha$ and
$k_{1} = k_{0} n_{rel}$ denote the wavenumber interior to that
sphere. A plane wave is incident on the sphere in direction
$\hat{\boldsymbol{\imath}}$, which we denote by $U_{i}$. The scattered
field exterior to the sphere is
\begin{equation}
  U_{s}(R,\hat{\boldsymbol{o}}) = \sum_{n = 0}^{\infty} a_{n}
  h_{n}^{(1)}(k_{0} R) P_{n}(\hat{\boldsymbol{\imath}} \cdot
  \hat{\boldsymbol{o}}),
\end{equation}
with $h_{n}^{(1)}$ denoting the spherical Hankel function of the first
kind with order $n$, and $P_{n}$ denoting the Legendre polynomial of
order $n$. The field interior to the sphere is
\begin{equation}
  U_{int}(R,\hat{\boldsymbol{o}}) = \sum_{n = 0}^{\infty} b_{n}
  j_{n}(k_{0} R) P_{n}(\hat{\boldsymbol{\imath}} \cdot
  \hat{\boldsymbol{o}}),
\end{equation}
with $j_{n}$ denoting the spherical Bessel function. We determine the
expansion coefficients $a_{n}$ and $b_{n}$ by requiring
that $U_{i} + U_{s} = U_{int}$ and
$\partial_{r} U_{i} + \partial_{r} U_{s} = \partial_{r} U_{int}$ on
$R = \alpha$. We compute a numerical approximation by truncating the
series at $n = 32$ and making use of the orthogonal properties of
Legendre polynomials.

Using the asymptotic behavior $h_{n}^{(1)}(z) \sim \mathrm{i}^{-n-1}
z^{-1} e^{\mathrm{i} z}$ as $z \to \infty$, we find that
\begin{equation}
  U_{s}(R,\hat{\boldsymbol{o}}) \sim \left[ \frac{1}{k_{0}} \sum_{n =
      0}^{\infty} a_{n} \mathrm{i}^{-n-1}
    P_{n}(\hat{\boldsymbol{\imath}} \cdot \hat{\boldsymbol{o}})
  \right] \frac{e^{\mathrm{i} k_{0} R}}{R},
\end{equation}
in the asymptotic limit, $R \to \infty$. The bracketed term in the
expression above gives the scattering amplitude $f$. Next, we use
$P_{n}(-1) = (-1)^{n}$ to determine that
\begin{equation}
  f(\hat{\boldsymbol{\imath}},-\hat{\boldsymbol{\imath}}) =
  \frac{1}{k_{0}} \sum_{n = 0}^{\infty} (-1)^{n} \mathrm{i}^{-n-1}
  a_{n}.
\end{equation}
We approximate this scattering amplitude by truncating the series as
we have done to determine the expansion coefficients. That result is
used as our frequency dependent reflectivity. We compute different
reflectivities by specifying different values of the radius $\alpha$
and the reflective index $n_{rel}$.

\section*{Acknowlegments}

This work was supported by the AFOSR Grant FA9550-21-1-0196. A.~D.~Kim
also acknowledges support by NSF Grant DMS-1840265).



\end{document}